\documentclass[runningheads]{llncs}
\usepackage{graphicx}
\usepackage[table]{xcolor}
\usepackage{multirow}
\usepackage{booktabs}
\newcommand{\ra}[1]{\renewcommand{\arraystretch}{#1}}

\begin{document}
\title{Linguistic and Gender Variation in Speech Emotion Recognition using Spectral Features\thanks{This publication has emanated from research supported in part by a Grant from Science Foundation Ireland under Grant number 18/CRT/6222}}
\titlerunning{Linguistic \& Gender Variation in SER using Spectral Features}
%
\author{Zachary Dair, Ryan Donovan, \and
Ruairi O'Reilly \orcidID{0000-0001-7990-3461}}
\authorrunning{Z. Dair et al.}
%
\institute{Munster Technological University, Cork, Ireland \\
\email{zachary.dair@mycit.ie, brendan.donovan@mycit.ie, ruairi.orielly@mtu.ie}\\
\url{www.mtu.ie}}
\maketitle              
\begin{abstract}
\vspace{-2em}
This work explores the effect of gender and linguistic-based vocal variations on the accuracy of emotive expression classification. Emotive expressions are considered from the perspective of spectral features in speech (Mel-frequency Cepstral Coefficient, Melspectrogram, Spectral Contrast). Emotions are considered from the perspective of Basic Emotion Theory. A convolutional neural network is utilised to classify emotive expressions in emotive audio datasets in English, German, and Italian. Vocal variations for spectral features assessed by (i) a comparative analysis identifying suitable spectral features, (ii) the classification performance for mono, multi and cross-lingual emotive data and (iii) an empirical evaluation of a machine learning model to assess the effects of gender and linguistic variation on classification accuracy. The results showed that spectral features provide a potential avenue for increasing emotive expression classification. Additionally, the accuracy of emotive expression classification was high within mono and cross-lingual emotive data, but poor in multi-lingual data. Similarly, there were differences in classification accuracy between gender populations. These results demonstrate the importance of accounting for population differences to enable accurate speech emotion recognition. 


\vspace{-2em}
\end{abstract}
\section{Introduction}\vspace{-1em}

Speech emotion recognition (SER) is the classification of the emotional states of a speaker from speech. These emotional states can either be discrete experiences, such as the basic emotions (Anger, Disgust, Fear, Joy, Sadness, and Surprise), or attributes of emotional states (Arousal, Valence, Dominance). The ability to accurately classify these emotional states through SER enables the provision of tailored services that can adapt to the psychological needs of user groups. SER attempts to classify these emotional states via verbal and non-verbal components of speech. Verbal components of speech are the specific words a speaker says. Verbal SER involves classifying emotional expressions from the word choice or associations, syntax, use of colloquialisms or sarcasm concerning the intent of the speaker \cite{goddard2011semantic}. Non-verbal components of speech refer to how the speaker expresses their speech. Non-verbal SER involves classifying emotional expressions from the speaker's acoustic features (Pitch, Tone, Timbre, Number of pauses, Loudness, Speech Rate) while speaking \cite{hsu2021speech}.

Research has shown that non-verbal components provide valuable information, distinct from verbal content, in accurately classifying emotional expressions. For example, non-verbal features are capable of indicating emotional signals when the verbal content of the speech is neutral \cite{soleymani2017survey} \cite{sauter2010perceptual}. Many non-verbal features have been shown to exist across multiple languages, enabling generalisable analysis \cite{sauter2010perceptual} \cite{sauter2015emotional}. Additionally, non-verbal features have been shown to have a direct impact on emotion recognition processes in the brain, indicating they are vital to achieving accurate emotion classification \cite{schirmer2017temporal}. Overall, the larger amount of features available for non-verbal SER, in comparison to verbal SER, enables more accurate emotion classification. 

Traditional non-verbal SER\footnote{Hereonafter, non-verbal SER is referred to as SER as shorthand.} approaches recruit human observers, who are asked to classify emotions from non-verbal content collected from one or several speakers. While traditional approaches are useful for relatively small-scale datasets, as they demonstrate high levels of accuracy \cite{sauter_cross-cultural_2010} \cite{sauter2015emotional} \cite{cordaro2016voice}, they are impractical for dealing with large and continuously growing datasets that characterise modern human-computer interactions. In an attempt to scale the benefits of SER, research and industrial applications have been developed to conduct SER automatically. These applications range from open-source research tools (for example, \textit{Audeering}) to commercial tools (for example, \textit{Good Vibrations}, \textit{Vokaturi}, and \textit{deep affect API}) \cite{garcia2017emotion} \cite{eyben2015real}, all of which employ machine learning-based approaches. If SER can be accurately automated, it would facilitate increased efficacy in human-computer interactions, user modelling, and personalisation services.

Despite the potential benefits of automated SER, it has not been widely implemented in everyday life settings \cite{akcay_speech_2020}. This is despite reviews of the SER literature demonstrating that automated SER approaches perform similarly or even better than traditional approaches \cite{akcay_speech_2020}. One reason for the limited uptake of automated SER approaches is that they are not as generalisable and adaptable as human classifiers, who can use contextual information (for example, the gender of the speaker and the language of the speech) when classifying speech. These two factors, gender and language, have been shown to significantly affect the accuracy of automated SER approaches \cite{lausen2018gender} \cite{rajoo2016influences}. 

SER models that account for gender differences are more accurate than those that do not \cite{xu2018effects}. It is difficult, however, to accurately model for these differences. While there exists many features of speech that are significantly influenced by gender, there exists considerable overlap between males and females on these same features. This overlap makes it difficult to distinguish between the gender and linguistic features of emotions that reliably indicate emotions and those that do not. Concerning language, SER models trained on one language significantly decrease in performance when tested on another language. This contrasts to human observers in traditional SER, who are capable of detecting emotional expressions in speech in languages they do not speak \cite{cordaro2016voice} \cite{sauter_cross-cultural_2010} \cite{sauter2015emotional}.

One potential mechanism for improving the accuracy of SER across different gender and linguistic populations are spectral features. Spectral features of audio are computed by converting a time-based signal into the frequency domain. These features represent the distribution of energy across a frequency and the harmonic components of sound. Components such as pitch changes in the audio signal, can assist in gender differentiation, and voice quality, can be identified through voice level as either tense, harsh, breathy, or lax \cite{voice_quality_gobl}. 
Aggregating  spectral features' continuous, qualitative, and spectral data aids differentiation between genders and linguistic populations thereby enabling more generalisable and accurate emotive expression classifications.

This study assesses the effect of spectral features on the performance of automatic SER across different gender and linguistic populations by (i) conducting a comparative analysis of spectral features to identify a suitable feature set, (ii) comparing classification performance of SER approaches when using mono-lingual, multi-lingual, and cross-lingual emotive data, and (iii) empirically evaluating the performance of a machine learning (ML) model to assess the effect of gender and linguistic variations on SER classification accuracy.\vspace{-1em}
\section{Related Work}
\vspace{-1em}

Previous research has examined the differences between gender populations in acoustic features and their effect on SER performance. Several acoustic features have been described as ``gender-dependent''. Gender dependent in this case means that gender significantly influences the expression of that feature \cite{vogt2006improving}. Pitch is an example of a gender-dependent feature. Females score higher, on average, than males on pitch \cite{latinus2012discriminating}. This information, however, is not sufficient to enable generalisable SER classification, because there is considerable overlap between males and females on this feature \cite{latinus2012discriminating}. This is also the case for several other acoustic features described as ``gender-dependent''. 

Similarly, previous research has examined the impact of language in developing accurate SER models. Human observers can accurately classify emotions in their native language, and non-native languages \cite{cordaro2016voice}. This result has been replicated across various cultures, including indigenous tribes, demonstrating the generalisability of traditional SER \cite{sauter_cross-cultural_2010} \cite{sauter2015emotional}. Similarly, automated SER models achieve a high level of accuracy in mono-lingual settings, where the models are trained and evaluated in the same language. The performance of these models, however, significantly drop when they are tested across languages \cite{feraru2015cross} \cite{neumann2018cross}.

If ML models can differentiate between acoustic features that indicate emotional expressions for each gender and linguistic population reliably, then one should expect generalisability of SER performance between both populations \cite{xu2018effects}. To achieve this, spectral features that reliably indicate emotional experiences are first required. 

A survey \cite{el_ayadi_survey_2011} analysed features across 17 distinctive datasets. These datasets comprise 9 languages, both male and female speakers, professional and non-professional acting sessions, recorded call centre conversations and speech recorded under simulated and naturally occurring stress situations. 

A pool of acoustic features and SER were analysed in \cite{el_ayadi_survey_2011} and \cite{akcay_speech_2020}. Which can be divided into several categories (Continuous, Qualitative, Spectral and Non-linear Teager Energy Operated (TEO). The results of both reviews indicated that Continuous and Spectral features were related to emotional expressions in speech. A weaker relationship was found between emotional expressions and Qualitative and TEO features.

To automatically classify relevant spectral features, a ML methodology is proposed in \cite{issa_speech_2020}. The methodology uses a convolutional neural network (CNN) and SER analysis. The approach consists of using spectral features such as Mel-frequency Cepstral Coefficients (MFCCs), Mel-scaled spectrogram, Chromagram, Spectral contrast and Tonnetz representation extracted from emotive audio belonging to three distinct datasets. This work, however, focused on a combined gender classification of emotive expression, inhibits the analysis of vocal variations between gender. This study extends this work to by analysing the effect of gender and linguistic populations on the emotive expression classification of spectral features. \vspace{-1em}

\section{Methodology}\vspace{-1em}
To assess the effect of spectral features for emotive expression classification across different gender and linguistic populations a series of experiments were conducted. Which required the identification of a suitable set of gender and/or language-dependent spectral features. The classification was enabled through the usage of the CNN described in \ref{cnn_section}, the performance of this CNN will be compared across experiments that account for linguistic and gender differences and experiments that do not account for these differences. The gender populations are male and female and the linguistic populations are English, German, and Italian.


Discrete emotional models identify several distinct emotions to be classified. This study uses Basic Emotion Theory (BET), as the discrete emotional model for SER. In BET the distinct emotions are Anger, Disgust, Fear, Joy, Sadness, and Surprise. These six emotions are considered basic as (i) they consistently correlate with psychological, behavioural, and neurophysiological activity \cite{ekman2004emotions}, which makes their objective measurement possible, and (ii) they interact to form more cognitively and culturally mediated emotions, such as shame or guilt \cite{ekman2004emotions}. The BET contrasts from dimensional models of emotions that focus on attributes of emotions (for example, arousal, valence). BET models have been extensively used in SER research as they capture a wider range of emotions and are intuitive to label in comparison to dimensional models \cite{akcay_speech_2020}.\vspace{-1em} 

\subsection{Emotive Speech Data}

Three distinct emotive audio datasets were used (see Table~\ref{tab1}) to enable a comparative analysis of variations between the data. These variations can originate from many factors such as the gathering method, emotions exhibited, language, speaker gender, or sampling rate.

\begin{table*}[ht]\centering
\ra{1.1}
    \caption{Details of the emotive expression datasets utilised.}\label{tab1}
    \begin{footnotesize}
    \begin{tabular}{lllllll}
        \toprule
        Datasets & \phantom{a} & RAVDESS\cite{livingstone2018ryerson} & \phantom{a} & EMO-DB\cite{burkhardt2005database}& \phantom{a} & EMOVO\cite{costantini_emovo_nodate} \\
        \midrule
        Population && 12M/12F && 5M/5F && 3M/3F \\
        Actor Ages && 21-33 && 21-35 && 23-30\\
        Professional Actors && Yes && Yes && Yes\\
        Language && English && German && Italian \\
        Emotions && A, F, J, S, D, Sur, N \footnote{Calm and Neutral combined} && A, F, J, B, S, D, N\footnote{Surprise absent, inclusion of Boredom} && A, F, J, S, D, Sur, N \\
        Total Instances && 1440 && 535 && 588 \\
        No. Unique Utterences && 2 && 10 && 14 \\
        Sampling Rate && 48kHz && 48kHz(16kHz) && 48kHz \\
        Human Reported Acc && 0.62 && 0.86 && 0.80 \\
        Reported Acc && 0.71\cite{issa_speech_2020} 0.74\cite{issc_paper} && 0.82\cite{issa_speech_2020} && 0.73\cite{ANCILIN2021108046} \\
        \bottomrule
    \end{tabular}
    \end{footnotesize}
\end{table*}

{\bfseries RAVDESS} contains emotional speech data constituting statements and songs in English. For the proposed work, only speech samples are considered. Participants uttered two neutral statements across multiple trials. For each trial, participants were asked to utter the statement in a manner that conveyed one of the six basic emotions. The statements were controlled to ensure equal levels of syllables, word frequency, and familiarity to the speaker. 

{\bfseries EMO-DB} contains emotional speech data constituting statements in German. Participants uttered ten statements, comprised of everyday language and syntactic structure, in various lengths to simulate natural speech. Each utterance was evaluated with regards to recognisability and naturalness of emotions exhibited. The emotive reaction of surprise is not considered. 

{\bfseries EMOVO} contains emotional speech data constituting statements in Italian. The participants uttered fourteen distinct semantically neutral statements. Each conveys a basic emotion and is spoken naturally. An important consideration in the creation of this dataset was the presence of all phonemes of the Italian language and a balanced presence of voiced and unvoiced consonants.\vspace{-1em}

\subsection{Feature Selection}\label{featureSelection}

In order to identify spectral features indicative of emotive expression, feature selection is performed. Which identifies information such as pitch, energy, voice levels and energy distribution for classification. Spectral features were extracted, using the audio analysis library Librosa \cite{brian_mcfee-proc-scipy-2015}. 

These features are as follows: Mel-Frequency Cepstral Coefficients (MFCC) which represents the shape of a signal spectrum, and is achieved by collectively representing a Mel-frequency cepstrum per frame. Chroma Energy Normalized (CENS) which represents statistics indicative of normalised values quantifying tempo, articulation, and pitch deviations. Zero-Crossing Rate (ZCR) which represents the rate of change in the audio signal from positive to negative or negative to positive, through zero. Chromagram which represents a transformed signal spectrum built using the chromatic scale capturing the harmonic and melodic characteristics. Melspectrogram which represents a spectrogram where the frequency is converted from a linear scale to a Mel-scale, resulting in a spectrogram reflective of how humans perceive frequencies, capturing  the amplitude of the signal. Spectral Contrast which represents the energy contrast computed by comparing the peak energy and valley energy in each band converted from spectrogram frames. Tonnetz which represents the tonal features in a 6-dimensional format, capturing traditional harmonic relationships. (perfect fifth, minor third and major third).

In order to identify a set of spectral features that captures sufficient emotive data, a classifier is trained on these features individually before being trained on permutations of the features across the datasets to identify the best performing feature set.\vspace{-1em}

\subsection{Linguistic Variation}
\vspace{-1em}
To evaluate the accuracy of a SER model in relation to linguistic variation three experiments are conducted across three languages (English, German, Italian) as follows:
\vspace{-1em}
\begin{enumerate}
    \item Mono-Lingual - The CNN is used to classify spectral features indicative of emotive expression from each language independently. The results will be used as a baseline performance of the model's classification performance in a single language.
    \item Multi-Lingual - This experiment consists of three permutations. For each, the CNN is first trained on one of the chosen languages and then evaluated for SER accuracy against the remaining languages. The intent is to identify spectral features capacity to classify emotive expression independent of language.
    \item Cross-Lingual - The CNN is trained on an aggregation of all training data. This will provide insights into the performance of a CNN model when its training corpus contains multiple languages. The intent is to identify the generalisability of the chosen spectral features for emotive expression classification.
\end{enumerate}\vspace{-2em}

\subsection{Cross-Gender Emotion}
\vspace{-1em}
To assess the effect of gender populations on SER, an experiment was conducted using the spectral features identified in Section.~\ref{featureSelection}, comprising of two steps. In step one, the data is classified using the six basic emotions. This initial classification provided a baseline accuracy where gender is not specified. In step two, the same dataset is classified with gender-emotion labels (for example, Male-Anger/Female-Anger; Male-Joy/Female-Joy). The intent is to evaluate if the CNN can identify gender-dependent acoustic elements from the spectral features.
\begin{figure}
\includegraphics[width=\textwidth]{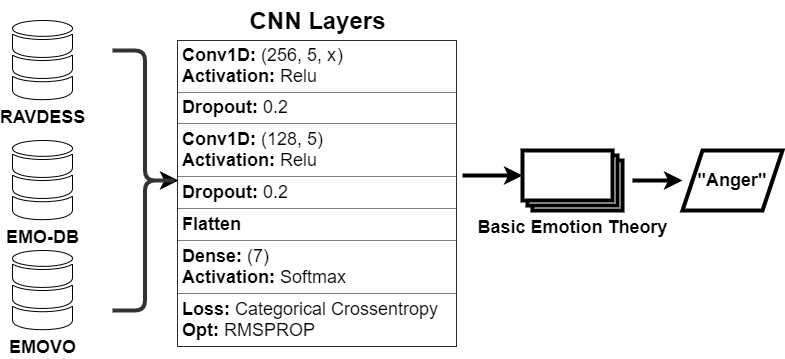}
\caption{An overview of the proposed approach, detailing the CNN layer structure.} \label{fig1}
\end{figure}
\vspace{-1em}
\subsection{Classification Of Emotive Speech}\label{cnn_section}
\vspace{-1em}
The architecture of the CNN used for classification is depicted in Figure \ref{fig1}. A CNN was re-implemented, due to the high emotive expression classification performance using spectral features as exhibited in \cite{issc_paper}. Optimal hyper-parameters were identified from a comparative analysis against related approaches. The model is trained on the extracted spectral features over 150 epochs using a batch size of 16 and 5-Fold validation is performed. During each epoch a portion of data is used to evaluate the model. The model performance is evaluated across the metrics precision, recall, F1-score for each of the six basic emotions. 

\vspace{-1em}
\section{Results}
\vspace{-1em}
\textbf{Feature extraction} - The comparative analysis of the spectral features in isolation provides insights into the performance on a per feature basis. MFCC, Melspectrogram and Spectral Contrast were the highest performing individual spectral features across the datasets. When combined these features formed a vector of 155 data-points, and the highest performing permutation of spectral features as denoted in Table.~\ref{exp_1_table}
\begin{table*}[ht]\centering
\ra{1.2}
    \caption{The classifiers F1 scores for individual spectral features and the top 4 combinations of spectral features. Results are listed in descending order based on the mean F1 score for single/multi feature classification across the datasets.}
    \label{exp_1_table}
    \centering
    \begin{footnotesize}     
    \begin{tabular}{@{}llllllll@{}}
        \toprule
         Spectral Features & \phantom{abc} & RAVDESS & EMO-DB & EMOVO & Mean\\
         \midrule
         MFCC && 0.69 & 0.71 & 0.71 & 0.70\\
         Melspectogram && 0.44 & 0.66 & 0.63 & 0.58\\
         Spectral Contrast && 0.39 & 0.41 & 0.45 & 0.42\\
         CENS && 0.31 & 0.44 & 0.34 & 0.36\\
         STFT && 0.22 & 0.42 & 0.37 & 0.34\\
         ZCR && 0.23 & 0.32 & 0.15 & 0.24\\
         Tonnetz && 0.22 & 0.27 & 0.20 & 0.23\\
         \midrule
         
         Spectral Feature Permutations && & & &\\ \midrule
         MFCC, Melspectogram, Spectral Contrast && 0.70  & 0.76  & 0.78. & 0.75\\
         Spectral Contrast, Melspectogram, MFCC && 0.68 & 0.76 & 0.75 & 0.73\\
         MFCC, STFT, Spectral Contrast && 0.67 & 0.70 & 0.69 & 0.69\\
         STFT, Melspectogram, Spectral Contrast, Tonnetz  && 0.51 & 0.67 & 0.60 & 0.59\\
         \midrule
         
         
         
         Additional Characteristics && & & &\\ \midrule
         Average Absoulte Amplitude (dB) && 0.07 & 0.62 & 0.17 & NA\\
         \bottomrule
    \end{tabular}     
    \end{footnotesize}
\end{table*}
\begin{table*}[ht]\centering
\ra{1.2}
\caption classification accuracy per language derived from 5-fold cross-validation, and train/test data specified. (\textit{X represents data from the corresponding column name)}\label{exp_2_table}
\begin{footnotesize}     
\begin{tabular}{@{}llllllllll@{}}
\toprule
Method & Train/Test & \phantom{abc} & English & \phantom{abc} & German & \phantom{abc} & Italian & \phantom{abc} & Combined\\
\midrule

Mono-Ling. & X/X && 0.70 && 0.76 && 0.78 && NA\\
Multi-Ling.-Eng. & RAVDESS/X &&  0.70 && 0.20 && 0.16 && NA\\
Multi-Ling.-Ger. & EMO-DB/X && 0.15 &&  0.76 && 0.16 && NA\\
Multi-Ling.-Ital. & EMOVO/X && 0.17 && 0.26 &&  0.78 && NA\\
Cross-Ling. & All/All && NA && NA && NA && 0.66\\


         \bottomrule
    \end{tabular}     \end{footnotesize}
\end{table*}

\textbf{Linguistic variation} - The results for emotive analysis in mono, multi and cross-lingual data are denoted in Table.~\ref{exp_2_table}, these highlight the considerations for SER across multiple languages. Mono-lingual and Cross-lingual approaches achieve high accuracies. A degradation in performance is experienced from multi-lingual approaches.
\begin{table*}[ht]\centering
\ra{1.1}
    \caption{The classifiers Precision (Prec.), Recall (Rec.) and F1 scores (F1) per BET emotion. The results demonstrate classification performance of male and female emotive expression independently and the F1 scores of combined gender emotive speech classification across each dataset.}\label{exp_gender_RAVDESS}
    \centering{
    \begin{footnotesize}     
    \begin{tabular}{@{}lllllllllllllll@{}}
        \toprule
        & \phantom{ab} & &\phantom{a} 
        & \multicolumn{3}{c}{Male} &\phantom{a} 
        & \multicolumn{3}{c}{Female} &\phantom{a} 
        & \multicolumn{3}{c}{Combined}
        \\
        \cmidrule{5-7} \cmidrule{9-11} \cmidrule{13-15} 
        Dataset & \phantom{a} & Emotion && Prec. & Rec. & F1
        && Prec. & Rec. & F1
        && Prec. & Rec. & F1 \\ \cmidrule{3-15} 
        
        RAVDESS&& Anger && 0.76 & 0.76 & 0.76 && 0.76 & 0.84 & 0.80 && 0.75 & 0.68 & 0.71 \\
        EMO-DB&& Anger && 0.90 & 1.00 & 0.95 && 0.78 & 0.78 & 0.78 && 0.92 & 0.86 & 0.89 \\
        EMOVO&& Anger && 0.90 & 0.90 & 0.90 && 1.00 & 0.92 & 0.96 && 0.86 & 0.92 & 0.89 \\ \cmidrule{3-15} 
        RAVDESS&& Joy && 0.57 & 0.44 & 0.50 && 0.73 & 0.73 & 0.73 && 0.86 & 0.62 & 0.72 \\
        EMO-DB&& Joy && 0.75 & 0.75 & 0.75 && 0.46 & 0.67 & 0.55 && 0.65 & 0.87 & 0.74 \\
        EMOVO&& Joy && 0.83 & 0.77 & 0.80 && 0.60 & 0.60 & 0.60 && 0.88 & 0.64 & 0.74 \\ \cmidrule{3-15} 
        RAVDESS&& Fear && 0.59 & 0.59 & 0.59 && 0.85 & 0.85 & 0.85 && 0.83 & 0.71 & 0.76 \\
        EMO-DB&& Fear && 1.00 & 0.75 & 0.86 && 0.80 & 1.00 & 0.89 && 0.67 & 0.40 & 0.50 \\
        EMOVO&& Fear && 0.67 & 0.80 & 0.73 && 0.71 & 1.00 & 0.83 && 0.64 & 0.69 & 0.67 \\ \cmidrule{3-15} 
        RAVDESS&& Sadness && 0.50 & 0.73 & 0.59 && 0.81 & 0.57 & 0.67 && 0.49 & 0.68 & 0.57 \\
        EMO-DB&& Sadness && 0.57 & 0.80 & 0.67 && 1.00 & 1.00 & 1.00 && 0.53 & 1.00 & 0.69 \\
        EMOVO&& Sadness && 1.00 & 1.00 & 1.00 && 0.88 & 1.00 & 0.93 && 0.94 & 0.85 & 0.89 \\ \cmidrule{3-15} 
        RAVDESS&& Surprise && 0.67 & 0.48 & 0.56 && 0.68 & 0.79 & 0.73 && 0.62 & 0.85 & 0.72 \\
        EMOVO&& Surprise && 0.64 & 0.70 & 0.67 && 0.56 & 0.45 & 0.50 && 0.59 & 0.84 & 0.70 \\ \cmidrule{3-15} 
        RAVDESS&& Disgust && 0.60 & 0.80 & 0.69 && 0.73 & 0.84 & 0.78 && 0.77 & 0.52 & 0.62 \\ 
        EMO-DB&& Disgust && 0.67 & 0.50 & 0.57 && 0.88 & 0.70 & 0.78 && 1.00 & 0.80 & 0.89 \\
        EMOVO&& Disgust && 0.75 & 0.50 & 0.60 && 0.86 & 0.86 & 0.86 && 0.64 & 0.70 & 0.67 \\ \cmidrule{3-15} 
        RAVDESS&& Neutral && 0.89 & 0.84 & 0.86 && 0.89 & 0.86 & 0.88 && 0.76 & 0.84 & 0.80 \\
        EMO-DB&& Neutral && 0.67 & 0.67 & 0.67 && 1.00 & 0.50 & 0.67 && 0.83 & 1.00 & 0.91 \\
        EMOVO&& Neutral && 0.86 & 1.00 & 0.92 && 0.83 & 0.83 & 0.83 && 1.00 & 0.86 & 0.92 \\  
        
        \bottomrule
    \end{tabular}     \end{footnotesize}
    }
\end{table*}

\textbf{Cross-Gender Emotion} - Table.~\ref{exp_gender_RAVDESS} identifies discrepancies in the emotive classification performance, across both gender and language. This indicates a degree of vocal variance stemming from the population differences, highlighting considerations for gender-specific SER approaches.
\vspace{-1em}
\section{Discussion}
\vspace{-1em}
\textbf{Feature extraction} - There are several notable results from the feature extraction stage that are worth discussion. Firstly, the results of the comparative analysis showed that the MFCC feature enabled the highest emotion recognition accuracy within each dataset. Demonstrating the importance of modelling for phonetic properties, found within the speech signal shape, in enabling accurate emotion classification across languages. Secondly, the performance of the Melspectrogram feature varied across the datasets. Melspectrogram performed highly on EMO-DB and EMOVO and poorly on RAVDESS. This inconsistency was caused by the significantly lower amplitude within the audio data of RAVDESS compared to EMO-DB and EMOVO denoted in Table.~\ref{exp_1_table}. Participants in the RAVDESS dataset were given instructions to exhibit emotions with varying intensity normal and strong, additionally, post-processing procedures are likely the cause for the lower level of amplitude. This has significant consequence for classification when amplitude is utilised as a measure of emotive expression. Thirdly, Spectral Contrast in isolation provides greater average accuracy than the remaining individual features. Therefore, the contrast between peak and valley energy is partially indicative of emotion in emotive speech. The high performance of these features in isolation indicates their potential suitability for emotive expression classification. 
Finally, the results showed that combining these three features increased accuracies for each data set. This demonstrates the importance of including such spectral features in SER classifications.

\textbf{Linguistic variation} - SER accuracy decreased significantly between the different types of analysis. The high performance on mono-lingual analysis indicates the importance of the selected spectral features for each language. The poor performance in multi-lingual analysis, however, indicates a lack of universality between spectral representations of emotions across the datasets. For example, there was significant variation in the amplitude and pitch range across the languages, as identified in \cite{mennen_et_al}. Additionally, these differences likely stemmed from differences in data collection (equipment, volume, actor-microphone distance) across the datasets, thereby decreasing the performance of the CNN in multi and cross-lingual analysis. This highlights the need for standardised recording procedures within SER research, and/or additional procedures to normalize linguistic variances.

\textbf{Cross-Gender Emotion} - There were substantial discrepancies in spectral representations between the gender populations. As a result, SER performance between the genders differed significantly across the six basic emotions. These differences were varied across languages, indicating that there is a generalisable effect of gender on SER performance when using spectral features, however this is impacted by linguistic variation. For example, high energy emotions (such as Anger and Joy) were more clearly detected in German-speaking males than in other linguistic or gender groups. The German language is characterised by higher amplitude, energy and harsher speech when articulated within males \cite{mennen_et_al}. This likely contributed to higher accuracy of detecting those emotions from German-speaking males. Disgust and Fear, in contrast, were more accurately classified from the speech of females across each language. This may have resulted from differences in vocals, particularly pitch, between the gender groups \cite{pepiot_voice_2015}. Additionally, female voices tend to articulate in a softer manner \cite{male_female_speech_comparison} conducive to representing softer, lower amplitude emotions.

These results likely contributed to the weaker performance of combined gender classification in comparison to gender-specific classifications of emotion. Combined approaches can account for variations between genders and languages, however, in certain cases, a specific gender may act as a limiting factor reducing the overall accuracy.

\textbf{Limitations} - The major limitations of this work concern sampling issues. Firstly, the sample size across the three datasets is small (N = 40). Since sample sizes can exaggerate the variances between populations, then the small sample size in this study may have exaggerated real differences in spectral features between gender and linguistic populations. Secondly, the datasets were comparing linguistic populations across an unequal number of speakers. Different sample sizes will have higher or lower ranges of variability. Comparing different sample sizes makes it difficult to determine whether the results stem from real between-group differences or are the result of a higher level of ``noisy'' data found in small sample sizes. These limitations damage the generalisability of the work. In a follow-up study, the sample size of the overall dataset will be increased and the number of speakers per linguistic group will be controlled.

\vspace{-1em}
\section{Conclusion}\vspace{-1em}

This work was an exploratory analysis of linguistic and gender variation on the emotive classification of spectral features. The results showed that features using the Mel-Scale and representations of amplitude and energy are important in accurate SER across different gender and linguistic populations. Additionally, the results showed that higher energy emotions such as Anger, Joy, Surprise were easier to identify originating from male voices in a high amplitude, harsh language such as German. Similarly lower energy emotions such as Disgust and Fear were easier to identify from female voices in each language. These observations highlight the importance of signal amplitude and energy when analysing emotion across gender and language.



The performance of emotive expression classification across mono, multi and cross-lingual data provides insights into linguistic variation of emotive audio. Mono-lingual approaches are suitable as baselines in comparison to multi and cross-lingual approaches. A linguistic variance between emotive expression represented by spectral features can be identified. To overcome these variances, it is recommended that cross-linguistic approaches, that combine languages for training, are implemented.




Future work should explore whether classification accuracy is affected by (i) model optimization in terms of structure and/or (ii) comprehensive hyper-parameter tuning (epochs, batch size, optimizers) using a grid search technique. Additionally, the relationships between linguistics, gender and emotion should be explored from a vocal, psychological and technical classification perspective, with a larger sample size to garner further insights into improving SER generalisability. 
%
%
%
%
\vspace{-1em}
\bibliographystyle{splncs04}
\bibliography{ref}

\end{document}